\documentclass[12pt,preprint]{emulateapj}

\newcommand{\be}{\begin{equation}}
\newcommand{\ee}{\end{equation}}
\newcommand{\ba}{\begin{eqnarray}}
\newcommand{\ea}{\end{eqnarray}}

\begin{document}

\title{Self-calibrating interloper bias in spectroscopic galaxy clustering surveys}


\author{Yan Gong$^{*1}$, Haitao Miao$^1$, Pengjie Zhang$^{*2,3,4}$, Xuelei Chen$^{1,5,6}$}

\affil{$^1$ National Astronomical Observatories, Chinese Academy of Sciences, Beijing 100101, China. Email:gongyan@bao.ac.cn}
\affil{$^2$ Department of Astronomy, Shanghai Jiao Tong University, Shanghai 200240, China. Email:zhangpj@sjtu.edu.cn}
\affil{$^3$ Shanghai Key Laboratory for Particle Physics and Cosmology, People’s Republic of China}
\affil{$^4$ Tsung-Dao Lee Institute, Shanghai 200240, People’s Republic of China}
\affil{$^5$ University of Chinese Academy of Sciences, Beijing 100049, People’s Republic of China}
\affil{$^6$ Centre for High Energy Physics, Peking University, Beijing 100871, Peoples Republic of China}

\begin{abstract}
Contamination of interloper galaxies due to misidentified emission lines can be a big issue in the spectroscopic galaxy clustering surveys, especially in the future high-precision observations. We propose a statistical method based on the cross correlations of the observational data itself between two redshift bins to efficiently reduce this effect, and it also can derive the interloper fraction $f_{\rm i}$ in a redshift bin with a high level of accuracy. The ratio of cross and auto angular correlation functions or power spectra between redshift bins are suggested to estimate $f_{\rm i}$, and the key equations are derived for theoretical discussion.  In order to explore and prove the feasibility and effectiveness of this method, we also run simulations, generate mock data, and perform cosmological constraints considering systematics based on the observation of China Space Station Telescope (CSST). We find that this method can effectively reduce the interloper effect, and accurately constrain the cosmological parameters for $f_{\rm i}<1\%\sim10\%$, which is suitable for most future surveys. This method also can be applied to other kinds of galaxy clustering surveys like line intensity mapping.
\end{abstract}


\maketitle

\section{Introduction}

Spectroscopic galaxy clustering surveys can illustrate the 3-dimensional (3-d) cosmic large-scale structure (LSS) of matter distribution. It is a powerful tool to explore the formation and evolution of the LSS and galaxies, and study the properties of the contents of the Universe, e.g. dark energy and dark matter. Several next-generation spectroscopic surveys are planned and will perform observations in the near future, such as the ground-based high-quality surveys Prime Focus Spectrograph \citep[PFS;][]{Tamura16}, Multi-Object Optical and Near-infrared Spectrograph \citep[MOONS;][]{Cirasuolo20, Maiolino20}, Dark Energy Spectroscopic Instrument \citep[DESI;][]{Levi13} and MegaMapper \citep{Schlegel19}, and the space-based slitless spectroscopic surveys $Euclid$ \citep{Laureijs11}, Wide-Field Infrared Survey Telescope (WFIRST) or Nancy Grace Roman Space Telescope \citep[RST;][]{Spergel13} and China Space Station Telescope \citep[CSST;][]{Zhan11,Zhan18,Cao18,Gong19}. 

Given that these spectroscopic surveys are dedicated to map the LSS and investigate the nature of dark energy and dark matter in unprecedented precision, the systematics need to be carefully handled and effectively reduced. One of important systematical effect is the contamination of interloper galaxies due to misidentified emission lines. One emission line at redshift $z_1$ can be recognized as another line at a different redshift $z_2$, if $\lambda_1(1+z_1) = \lambda_2(1+z_2)$, where $\lambda_1$ and $\lambda_2$ are the wavelengths of the two emission lines. Previous studies have found that, although this effect can be effectively suppressed in the surveys with high spectral resolution by using spectral energy distribution (SED) template fitting methods and considering secondary emission lines, it should be significant and needs to be seriously considered in the slitless spectroscopic surveys, such as RST, $Euclid$, and CSST \citep[e.g. see][]{Pullen16}. This effect can lead to considerable bias on the power spectrum, growth rate, and other important quantities,  when extracting the cosmological information from these spectroscopic galaxy clustering measurements.

Since it is difficult and time-consuming to eliminate this contamination in the data processing stage, we can try to deal with it statistically when analyzing the data. We propose a statistical method to extract the interloper fractions in redshift bins, that can effectively reduce its effect in the cosmological analysis. An equation set composed of the cross and auto angular galaxy correlation functions or power spectra of redshift bins is derived, which can be solved theoretically to obtain the interloper fraction. To check the feasibility of this method in practice, we also take the CSST spectroscopic survey as an example, and run numerical simulations to generate mock galaxy catalogs. The interloper fractions and uncertainties can be derived in different redshift bins from the simulations. We then apply these results to the cosmological constraints with systematics considered, and explore its effect when fitting the mock data of the CSST redshift-space distortion (RSD) measurements. Several interloper fraction cases are investigated, and the constraints on the cosmological parameters and interloper fractions are obtained. We can see that this method can effectively reduce the interloper bias in the cosmological constraints, and provide accurate results in a large interloper fraction range.
\\ \\
\section{theory}

We propose a statistical and self-calibration method which uses the cross and auto angular galaxy correlation functions or power spectra of the observational data itself to derive the interloper fraction in a redshift bin. First, we derive the equations of angular correlation functions or power spectra including interloper galaxies. The total angular galaxy overdensity in redshift bin $i$ is given by
\be
\delta_{\rm t}^i({\theta}) = \frac{n_{\rm t}^i(\theta)-\bar{n}_{\rm t}^i}{\bar{n}_{\rm t}^i},
\ee
where $n_{\rm t}(\theta)$ and $\bar{n}_{\rm t}$ are the total galaxy surface number density in $\theta\pm1/2\,\delta\theta$ and total mean density of the whole redshift bin, respectively. Considering interloper galaxies, we have $n_{\rm t}=n_{\rm r}+n_{\rm i}$, where $n_{\rm r}$ and $n_{\rm i}$ are the surface number densities of real galaxies in bin $i$ and interloper galaxies from other bins, respectively. In redshift bin $i$, assuming we only have one main interloper line from bin $j$, the fraction of interlopers is $f_{j\to i}=N_{\rm i}^{j\to i}/N_{\rm t}^i$, where $N_{\rm i}^{j\to i}$ and $N_{\rm t}^i$ are the number of interloper galaxies from bin $j$ and total galaxies in bin $i$, respectively. Then we can notice that $n^i_{\rm t}=n^i_{\rm i}/f_{j\to i}=n^i_{\rm r}/(1-f_{j\to i})$. Since galaxies in bin $i$ and $j$ can contaminate each other, and then for these two redshift bins we have 
\ba
\left\{\begin{array}{ll}
\delta_{\rm t}^i({\theta}) &= (1-f_{j\to i})\,\delta_{\rm r}^i(\theta) + f_{j\to i}\,\delta_{\rm i}^{j\to i}(\theta),  \\
\delta_{\rm t}^j({\theta}) &= (1-f_{i\to j})\,\delta_{\rm r}^j(\theta) + f_{i\to j}\,\delta_{\rm i}^{i\to j}(\theta).
\end{array}\right.
\ea
Here $\delta_{\rm r}(\theta)=[n_{\rm r}(\theta)-\bar{n}_{\rm r}]/\bar{n}_{\rm r}$ and $\delta_{\rm i}(\theta)=[n_{\rm i}(\theta)-\bar{n}_{\rm i}]/\bar{n}_{\rm i}$ are the surface overdensities of real and interloper galaxies in a redshift bin, respectively. Since we realize that $\delta_{\rm i}^{j\to i}(\theta)=\delta_{\rm r}^j(\theta)$ and 
$\delta_{\rm i}^{i\to j}(\theta)=\delta_{\rm r}^i(\theta)$, the auto and cross angular correlation functions $w(\theta)$ or power spectra $C_{\ell}$ for bin $i$ and $j$ can be calculated by
\ba \label{eq:Bt}
\left\{\begin{array}{ll}
B_{\rm t}^i &= (1-f_{j\to i})^2 B_{\rm r}^i + f_{j\to i}^2 B_{\rm r}^j,  \\
B_{\rm t}^j &= (1-f_{i\to j})^2 B_{\rm r}^j + f_{i\to j}^2 B_{\rm r}^i, \\
B_{\rm t}^{ij} &= f_{i\to j}(1-f_{j\to i}) B_{\rm r}^i + f_{j\to i}(1-f_{i\to j}) B_{\rm r}^j.
\end{array}\right.
\ea
Here $B=w(\theta)$ or $C_{\ell}$ is the angular correlation function or power spectrum, and $B^i$ and $B^{ij}$ denote the auto and cross correlation function or power spectrum, respectively. Since $B_{\rm t}(\theta\ {\rm or}\ \ell)$ can be measured in the observation, we have four unknown quantities in Eqs.~(\ref{eq:Bt}), i.e. $B_{\rm r}^i$, $B_{\rm r}^j$, $f_{j\to i}$, and $f_{i\to j}$. If we have $N$ bins in $\theta$ or $\ell$, there will be $3N$ measurable and $2N+2$ unknown quantities. This means that Eqs.~(\ref{eq:Bt}) can be solved theoretically when $N>2$.
Take the spectroscopic survey observing H$\alpha$6563$\rm \AA$ and [OIII]5007$\rm \AA$ lines as an example. The H$\alpha$ galaxies in redshift bin $0<z\lesssim0.3$ and $0.3\lesssim z\lesssim0.7$ can contaminate the [OIII] galaxies in $0.3\lesssim z\lesssim0.7$ and $0.7\lesssim z\lesssim1.2$, respectively, and vice versa. By using Eqs.~(\ref{eq:Bt}), in principle, we can derive the interloper fractions for these three large redshift bins. 

In the spectroscopic galaxy surveys, the 3-d galaxy power spectra $P_{\rm g}(k,z)$ at different redshifts with small redshift interval $\Delta z$ can be measured for analyzing cosmological information. Then we can make use of the derived fraction of interlopers as an average value in a redshift bin to estimate a set of total $P_{\rm g}(k,z)$ with redshift interval $\Delta z$ in this bin. Similar to the 2-d angular case, we can also derive the total 3-d power spectrum including interlopers \cite[see details in][]{Pullen16}. Note that the `projection' effect must be considered here, which can change the amplitude and scale of the 3-d interloper power spectrum \citep{Visbal10,Gong14,Gong17,Gong20,Lidz16,Pullen16}. Unlike the 2-d angular correlation case, for interloper galaxies from a redshift $z_j$ that contaminating galaxies at redshift $z_i$ in a 3-d volume, both the scales and the  volume elements will be changed when Fourier transforming the interloper correlation function to the power spectrum from $z_j$ to $z_i$. Consequently, additional factors need to be added to correct the scales and amplitude of the interloper power spectrum at $z_i$. As indicated in \cite{Pullen16},  the total 3-d galaxy power spectrum can be expressed as
\be \label{eq:Pt}
P_{\rm t}(k,z) = (1-{f}_{\rm i})^2 P_{\rm r}(k,z) + {f}_{\rm i}^2 A^2_{\perp}A_{\parallel}P_{\rm i}(k_{\rm i} ,z_{\rm i}).
\ee
Here $f_{\rm i}(z)=N_{\rm i}(z)/N_{\rm t}(z)$ is the fraction of interloper galaxies at $z$, and it can be approximately replaced by $\bar{f}^{\rm bin}_{\rm i}$, where $\bar{f}^{\rm bin}_{\rm i}$ is the average fraction of interloper galaxies in the redshift bin, which can be derived from the angular correlation function or power spectrum. $z_{\rm i}$ is the redshift of the interloper galaxies, $k_{\rm i}=\sqrt{A^2_{\perp}k^2_{\perp}+A^2_{\parallel}k^2_{\parallel}}$ is the wavenumber at $z_{\rm i}$, and $A_{\perp}=r(z)/r_{\rm i}(z_{\rm i})$ and $A_{\parallel}=y(z)/y(z_{\rm i})$, which represent the projection effect for interlopers on scales and amplitude of the power spectrum. Here $r$ is the comoving distance, and $y(z)={\rm d}r/{\rm d}\nu=\lambda(1+z)^2/H(z)$, where $\lambda$ is the rest-frame wavelength of the emission line, and $H(z)$ is the Hubble parameter. 

In Eq.~(\ref{eq:Pt}), we can find that the contribution of the interloper term will be insignificant when $z_{\rm i}\gg z$, especially for ${f}_{\rm i}\ll 1$ which should be the case in the future spectroscopic galaxy surveys. Hence, we can probably neglect the interloper galaxies from higher redshifts, that will not significantly affect the extraction of cosmological information, e.g. constraint on cosmological parameters. Bearing this in mind, we can further reduce Eqs.~(\ref{eq:Bt}) to a simpler form as
\be
\bar{f}^{\rm bin}_{\rm i} = f_{j\to i}=\frac{B_{\rm t}^{ij}}{B_{\rm t}^{j}},
\ee
assuming $z_i>z_j$. This form could be more practical than Eqs.~(\ref{eq:Bt}) in real data analysis, considering that we actually cannot measure $B_{\rm t}$ precisely in all scales, which can lead to problems when solving Eqs.~(\ref{eq:Bt}) numerically. We also notice that, since $B_{\rm t}(\theta\ {\rm or}\ \ell)$ is scale-dependent, it is useful to find a suitable scale range, and derive an average interloper fraction over this range to represent $f_{j\to i}$ or $\bar{f}_{\rm i}^{\rm bin}$. We run simulations to check the feasibility of our method and try to find this scale range as a reference in the following discussion.

\section{simulation}

\begin{figure}[t]
\includegraphics[scale = 0.33]{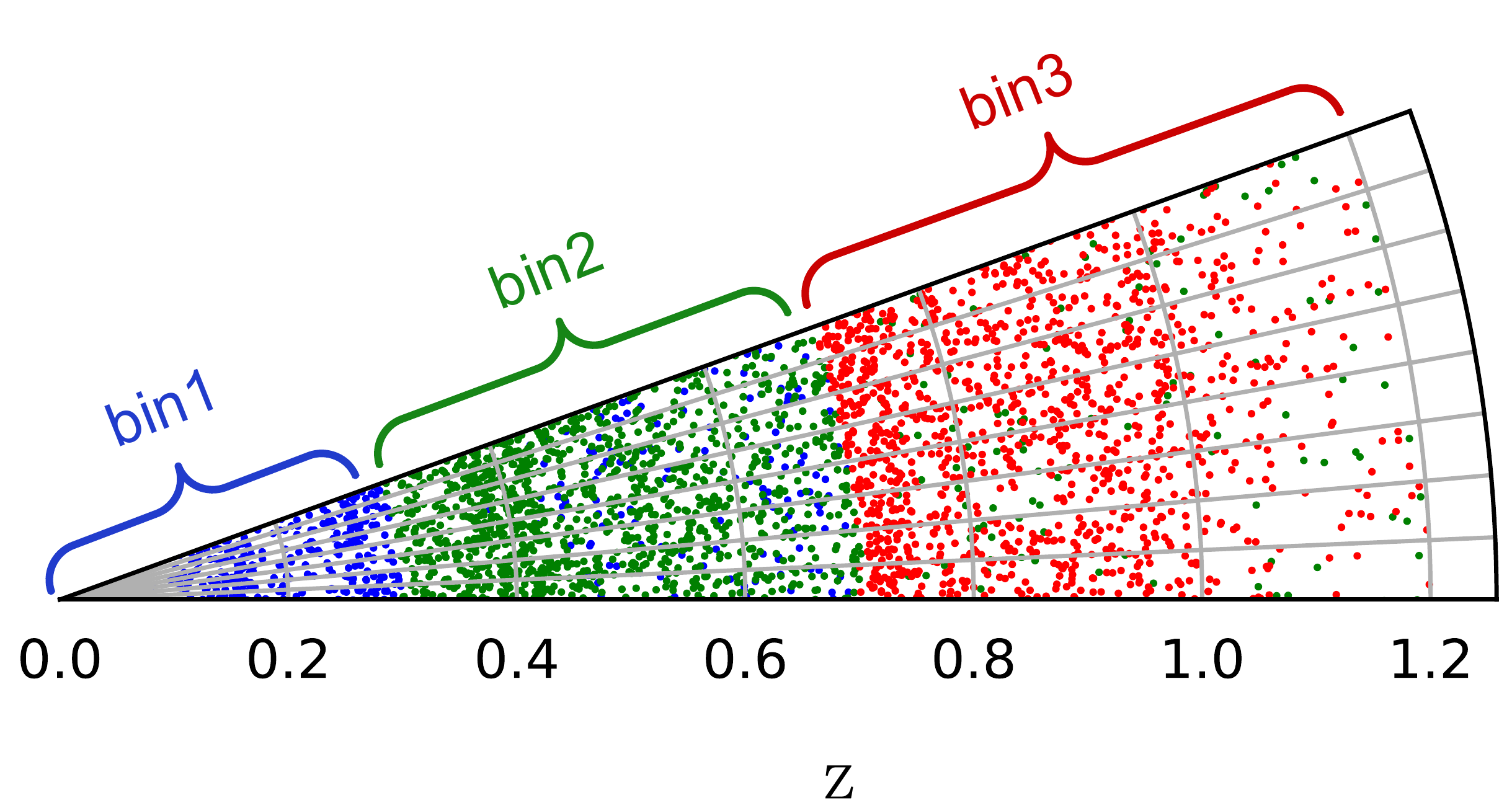}
\caption{The simulated lightcone for the CSST H$\alpha$ and [OIII] ELG mock catalog with H$\alpha$ interloper galaxies from lower redshifts. Only a small patch of sky and $\sim$10\% of the whole sample are shown here. We divide the light cone into three redshift bins, i.e. $0\le z\le0.28$ (bin1 with blue dots), $0.3< z\le0.68$ (bin2 with green dots), and $0.7< z\le1.2$ (bin3 with red dots), to estimate the auto and cross angular correlation functions of these bins. The contamination of [OIII] galaxies from higher redshift bins are neglected here.}
\label{fig:lc}
\end{figure}

\begin{figure*}
\centerline{
\includegraphics[scale=0.41]{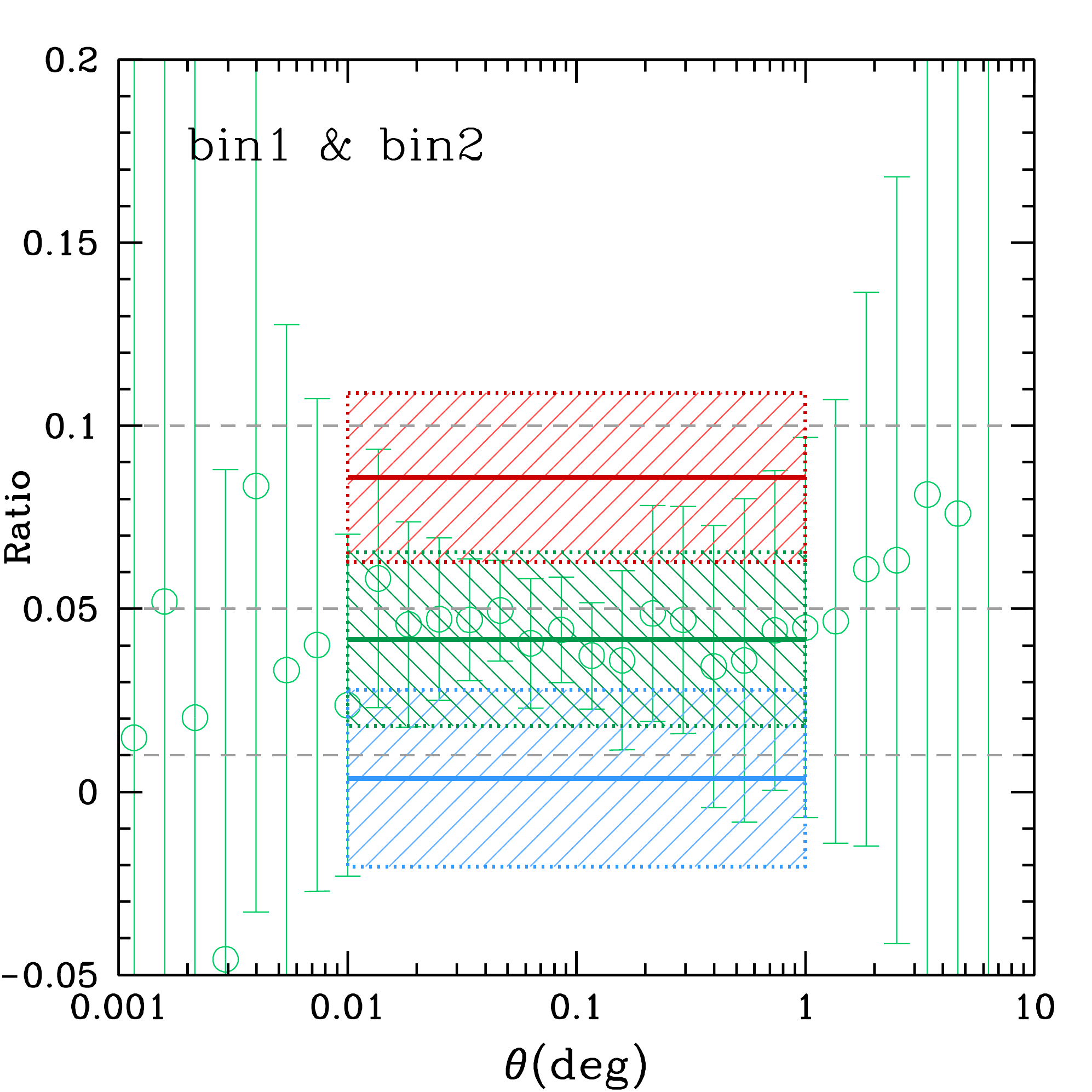}
\includegraphics[scale=0.41]{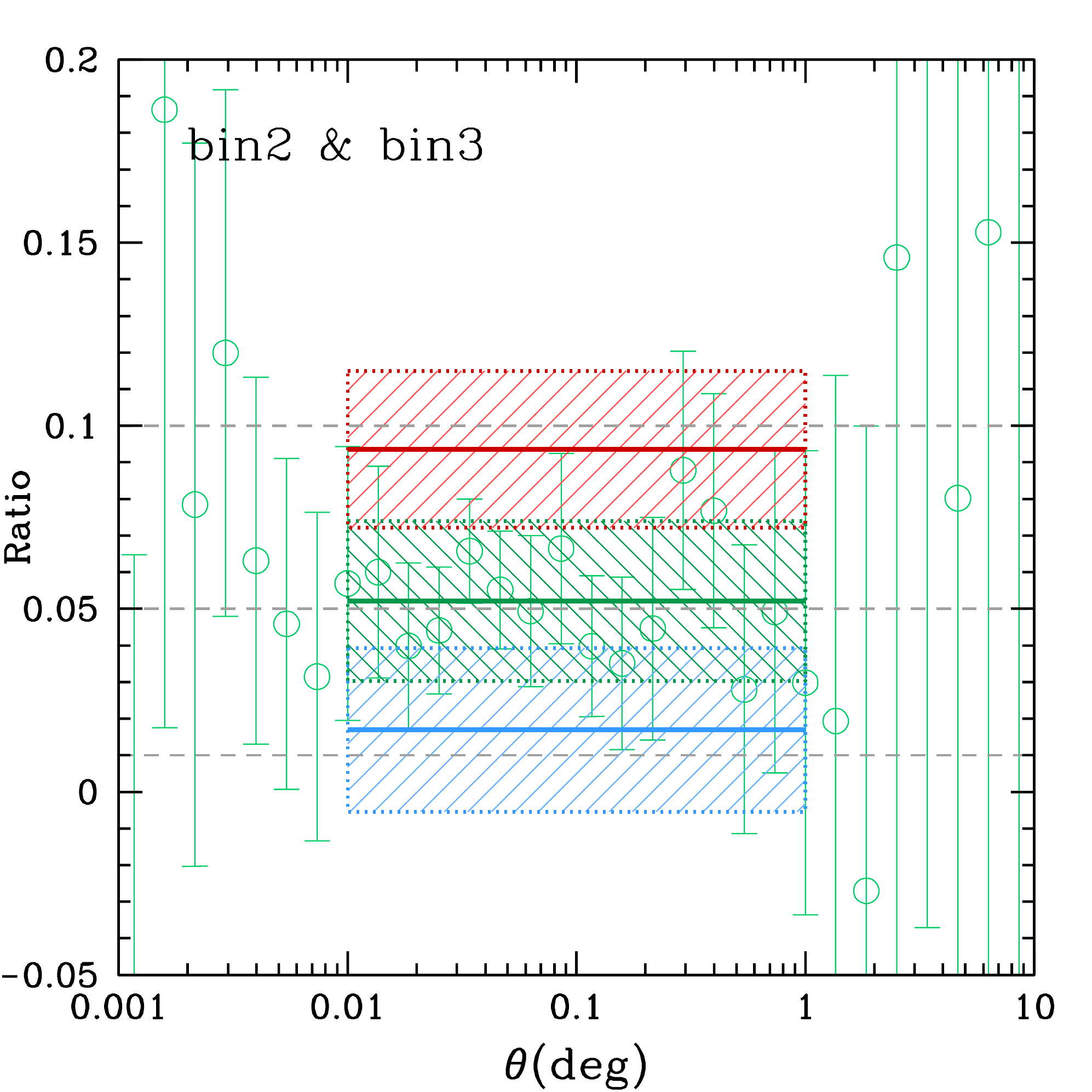}
}
\caption{The ratios between the cross and auto correlation functions. The left panel shows the result for bin1 and bin2, and right panel for the bin2 and bin3. As an example, the green data points with error bars are the ratios of the cross and auto correlation functions $w_{\rm t}^{ij}(\theta)/w_{\rm t}^{j}(\theta)$ assuming $\bar{f}_{\rm i}^{\rm bin}=5\%$. The blue, green, and red lines and boxes are the estimator $\bar{r}_{\rm t}$ we propose and errors $\Delta \bar{r}_{\rm t}$ over $\theta=10^{-2}$ to 1 deg by assuming $\bar{f}_{\rm i}^{\rm bin}=1\%$, $5\%$, and $10\%$ (the gray dashed lines), respectively.}
\label{fig:ratio}
\end{figure*}

We select the CSST spectroscopic galaxy survey as an example to run simulations \citep{Zhan11,Zhan18,Cao18,Gong19}. The CSST is a 2m aperture space telescope, which will be launched in 2024. It can simultaneously perform photometric imaging and slitless grating spectroscopic surveys covering 17,500 deg$^2$ in about 10 years. It has three spectroscopic bands, i.e. $GU$, $GV$, and $GI$, with wavelength coverage from $\sim$250 to $\sim$1100 nm and magnitude limit $\sim23$ AB mag for 5$\sigma$ point sources. The galaxies emitting H$\alpha$ and [OIII] lines are the main targets in the CSST spectroscopic survey, and the redshift distribution can extend to $z\gtrsim 1.5$  with a peak around $z\simeq0.3$. As estimated in previous works, the surface number density of  emission line galaxy (ELG) can reach $2\sim3$ arcmin$^{-2}$, and totally more than one hundred million ELGs can be measured in the CSST spectroscopic survey \citep{Gong19}.

We follow the CSST ELG redshift distribution and number density to simulate the mock catalogs in a lightcone covering 14,400 deg$^2$ and ranging from $z=0$ to 1.2. This area is actually large enough to represent the CSST survey, since it is expected that only $\sim$15,000 deg$^2$ can be left in the CSST survey, after masking badly measured area with image defects, reflections, ghosts, etc.\citep{Gong19}. We divide the lightcone into three redshift bins, i.e. $0\le z\le0.28$, $0.3< z\le0.68$, and $0.7< z\le1.2$, to explore the case for the H$\alpha$ and [OIII] emission line galaxies, which form the main sample in the CSST spectroscopic survey. The code L-PICOLA is used here to generate the lightcone \citep{LPICOLA}, which is an efficient and accurate parallel implementation of the COLA \citep{Tassev13}. The lightcone is generated by a simulation with a box size $1024\, h^{-1}\, \rm{Mpc}$ and $1024^3$ number of particles, and we start the simulation at redshift $z=9$. The L-PICOLA can directly perform lightcone simulations, and could replicate the box to reach the required redshift during runtime. We set 30 snapshots in each redshift bin, and totally 90 snapshots are used to construct the lightcone.

The ROCKSTAR is then used as a halo finder to build up the mock galaxy catalog with halo mass $M_{\rm h}>10^{10}\ M_{\sun}/h$  \citep{ROCKSTAR}. Since our main purpose here is attempting to find a suitable scale range as a reference for deriving $\bar{f}_{\rm i}^{\rm bin}$ from the ratio of galaxy angular correlation functions, for simplicity and considering the halo occupation distribution (HOD) model, we assume that only the halos with mass $M_{\rm h}>10^{10}\ M_{\sun}/h$ can host galaxies, and the satellite galaxies in a halo are ignored in our analysis. This assumption can be polished in the future works for further studies.

In Figure~\ref{fig:lc}, we show a small patch of the simulated lightcone of the CSST mock galaxy catalog for the spectroscopic survey. As discussed in the last section, we only consider the H$\alpha$ galaxies from lower redshift bins as interlopers (blue and green dots in bin2 and bin3, respectively, in Figure~\ref{fig:lc}), which are misidentified as [OIII] galaxies at higher redshift. The contamination of the bin1 is negligible, since the signal to noise ratios (SNRs) of the measured emission lines are high for the CSST spectroscopic survey with SNR$>$3 at $z<0.3$ \citep{Zhou21}, and the misidentification rate should be low enough to can be ignored.

We estimate the angular correlation function using the estimator \citep{Landy93}
\be 
w(\theta) = \frac{\langle DD\rangle - 2\langle DR\rangle + \langle RR\rangle}{\langle RR\rangle},
\ee
where $DD$, $DR$, and $RR$ are the pair counts of the data-data, data-random, and random-random points, respectively, in angular bins of $\theta$. The number of random objects we use is 80 times larger than that of the mock galaxies shown in the lightcone. The code CUTE is used in the estimate to obtain the correlation functions \citep{CUTE}, and the jackknife method is adopted to derive the covariance matrices of the correlation functions. 

In Figure~\ref{fig:ratio}, we show the ratios of cross and auto correlation functions $w_{\rm t}^{ij}(\theta)/w_{\rm t}^{j}(\theta)$ for bin1 and bin2 (left panel), and bin2 and bin3 (right panel). Here we take $\bar{f}_{\rm i}^{\rm bin}=1\%$, 5$\%$, and 10$\%$ as examples, considering the evaluation of the interloper fraction after secondary line identification for the future surveys \citep[e.g. see][]{Pullen16}. As expected, the interloper fraction can be as large as $\sim 1\%-10\%$ after data processing in the future slitless spectroscopic galaxy surveys. We can find that $w_{\rm t}^{ij}(\theta)/w_{\rm t}^{j}(\theta)$ (green dots with error bars) is in a good agreement with the assumed $\bar{f}_{\rm i}^{\rm bin}$ (gray dashed lines), and they are basically consistent with each other in 1$\sigma$. The scale range between $\theta=10^{-2}$ to 1 deg can be a suitable range to derive an average value of $w_{\rm t}^{ij}(\theta)/w_{\rm t}^{j}(\theta)$ as an estimator of $\bar{f}_{\rm i}^{\rm bin}$. If considering the weight of errors at different scales, the average ratio can be estimated by a simple form
\be
\bar{r}_{\rm t} = \frac{\sum_N a_n w_{\rm t}^{ij}(\theta_n)}{\sum_N a_n w_{\rm t}^{j}(\theta_n)}.
\ee
Here $N$ is the number of correlation function data in a scale range, $a_n = 1/[\sigma^2_{ij}(\theta_n)+\sigma^2_{j}(\theta_n)]$, where $\sigma_{ij}$ and $\sigma_j$ are the errors of the cross and auto correlation functions, respectively. In Figure~\ref{fig:ratio}, we can see that $\bar{r}_{\rm t}$ (blue, green, and red thick lines and boxes) can correctly represent the value of $\bar{f}_{\rm i}^{\rm bin}$ within the error $\Delta \bar{r}_{\rm t}\lesssim 0.02$. $\Delta \bar{r}_{\rm t}$ can be derived by the average values of the errors of $w_{\rm t}^{ij}(\theta)/w_{\rm t}^{j}(\theta)$ between $\theta=10^{-2}$ to 1 deg. It seems that the accuracy is not sensitive to the assumed value of $\bar{f}_{\rm i}^{\rm bin}=1\%$, 5$\%$, and 10$\%$ (gray dashed lines). Besides, the angular power spectrum $C_{\ell}$ also can be used to derive $\bar{r}_{\rm t}$, and it will be more convenient especially in the theoretical estimation.

\section{cosmological constraint}

After obtaining the derived $\bar{f}_{\rm i}^{\rm bin}$ given by $\bar{r}_{\rm t}$ and uncertainty $\Delta \bar{r}_{\rm t}$ from the angular correlation functions or power spectra for a redshift bin, we can explore its effect on the cosmological constraints in 3-d spectroscopic galaxy clustering surveys. Since the simulation above is too simple to obtain reliable 3-d galaxy power spectra, we adopt theoretical predications in the following analysis. This also will provide more flexibility and is suitable for exploring different cases in the current work. First, we generate the mock data of the galaxy power spectrum. Here we consider the RSD effect, and the total galaxy power spectrum is given by \citep{Pullen16}
\ba
P_{\rm t}(k,\mu,z) &=& (1-f_{\rm i})^2 P_{\rm r}(k, \mu, z) \\ \nonumber
&+& f_{\rm i}^2 A^2_{\perp}A_{\parallel}P_{\rm i}(k_{\rm i}, \mu_{\rm i}, z_{\rm i}),
\ea
where $\mu=k_{\parallel}/k$, and $\mu_{\rm i}=k_{\parallel, \rm i}/k_{\rm i}$. Here we assume a redshift-dependent $f_{\rm i}(z)=f_{n}(1+z)$, since $f_{\rm i}$ should be larger at higher redshifts due to poorer measurements. $f_{n}$ is the interloper fraction at $z=0$, and we will check the results for $f_{n}=0.01$, 0.05, and 0.1 in this work. The fiducial average interloper fraction in a redshift bin $\bar{f}_{\rm i}^{\rm bin}$ then can be calculated by $f_{\rm i}(z)$. Assuming there is no peculiar velocity bias, we have
\be
P(k,\mu,z)=P_{\rm g}(k,z)(1+\beta \mu^2)^2 D(k, \mu, z) + P_{\rm shot}(z).
\ee
Here $P_{\rm g}(k,z)=b^2_{\rm g}(z) P_{\rm m}(k,z)$, where $b_{\rm g}$ is the galaxy bias, and $P_{\rm m}$ is the matter power spectrum. We assume $b_{\rm g}=b_0(1+z)^{b1}$, and set $b_0$=1 and $b_1$=1 in the fiducial model. $\beta=f/b_{\rm g}$, where $f={\rm d\, ln}D(a)/{\rm d\, ln}a$ is the growth rate, and $D(a)$ is the growth factor. $D(k,\mu,z)$ is the damping factor at small scales, which can be expressed as 
\be
D(k,\mu,z)={\rm exp}\,[-(k\mu\sigma_{\rm D})^2].
\ee
Here $\sigma_{\rm D}=\sqrt{\sigma^2_{\nu}+\sigma^2_R}$, where $\sigma_{\nu}=\sigma_{\nu0}/(1+z)$ denotes the velocity dispersion effect \citep{Scoccimarro04,Taruya10}, and $\sigma_R=c\, \sigma_z/H(z)$ is the smearing factor at the scales below the spectral resolution $R$ in spectroscopic surveys. We assume $\sigma_{\nu0}=7$ Mpc$/h$ and $\sigma_z=(1+z)\sigma_z^0$ with $\sigma_z^0=0.002$ in the CSST survey \citep{Gong19}. Note that $D(k,\mu,z)$ cannot affect the power spectrum at large scales significantly, especially for the linear scales that we are interested in. $P_{\rm shot}=1/n_{\rm g}(z)$ is the shot-noise power spectrum, and $n_{\rm g}(z)$ is the galaxy number density at $z$. We estimate $n_{\rm g}(z)$ based on the mock data of the CSST spectroscopic survey, which is derived from the zCOSMOS catalog \citep{Lily07,Lily09}, and more details can be found in \cite{Gong19}.

\begin{figure}[htb]
\includegraphics[scale = 0.41]{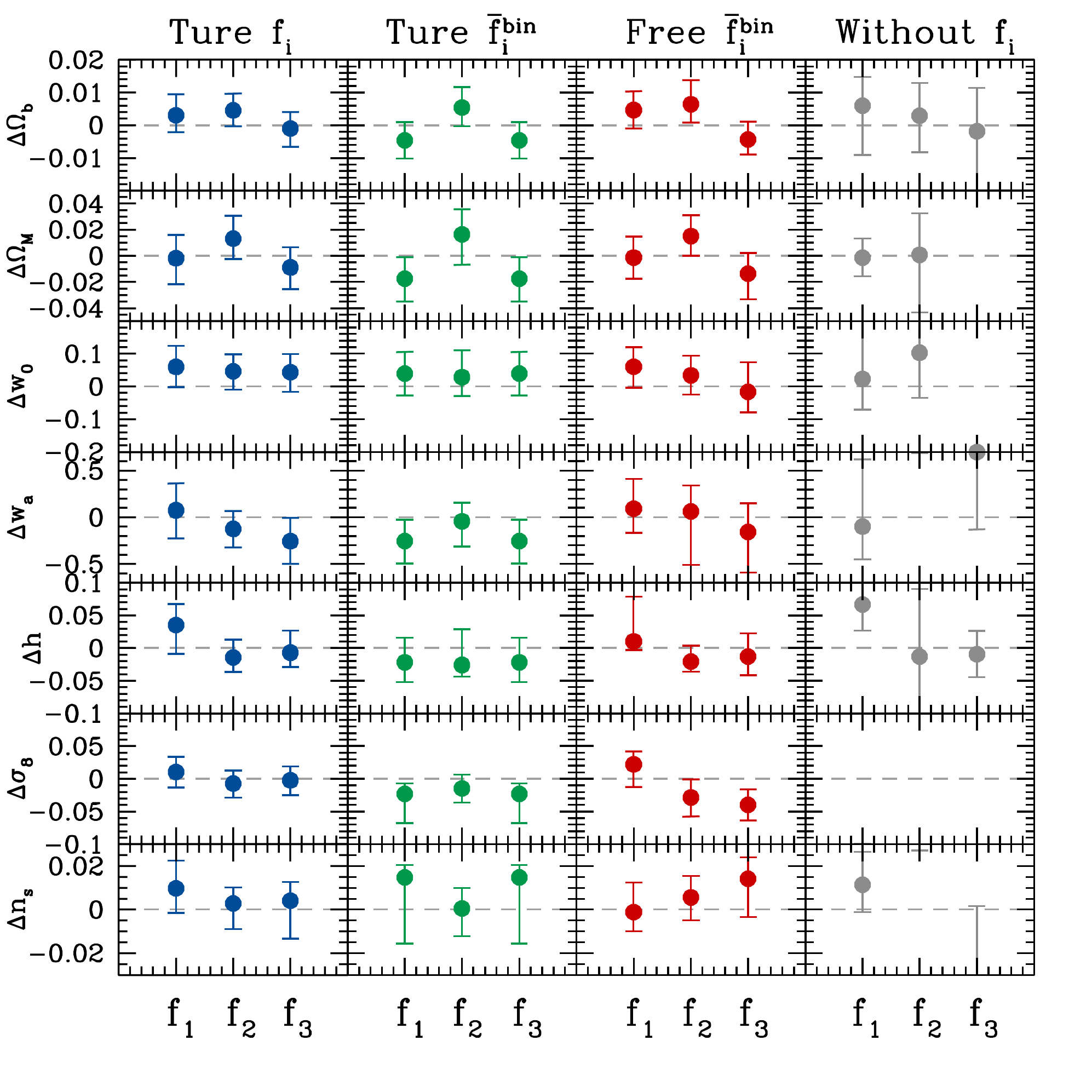}
\includegraphics[scale = 0.41]{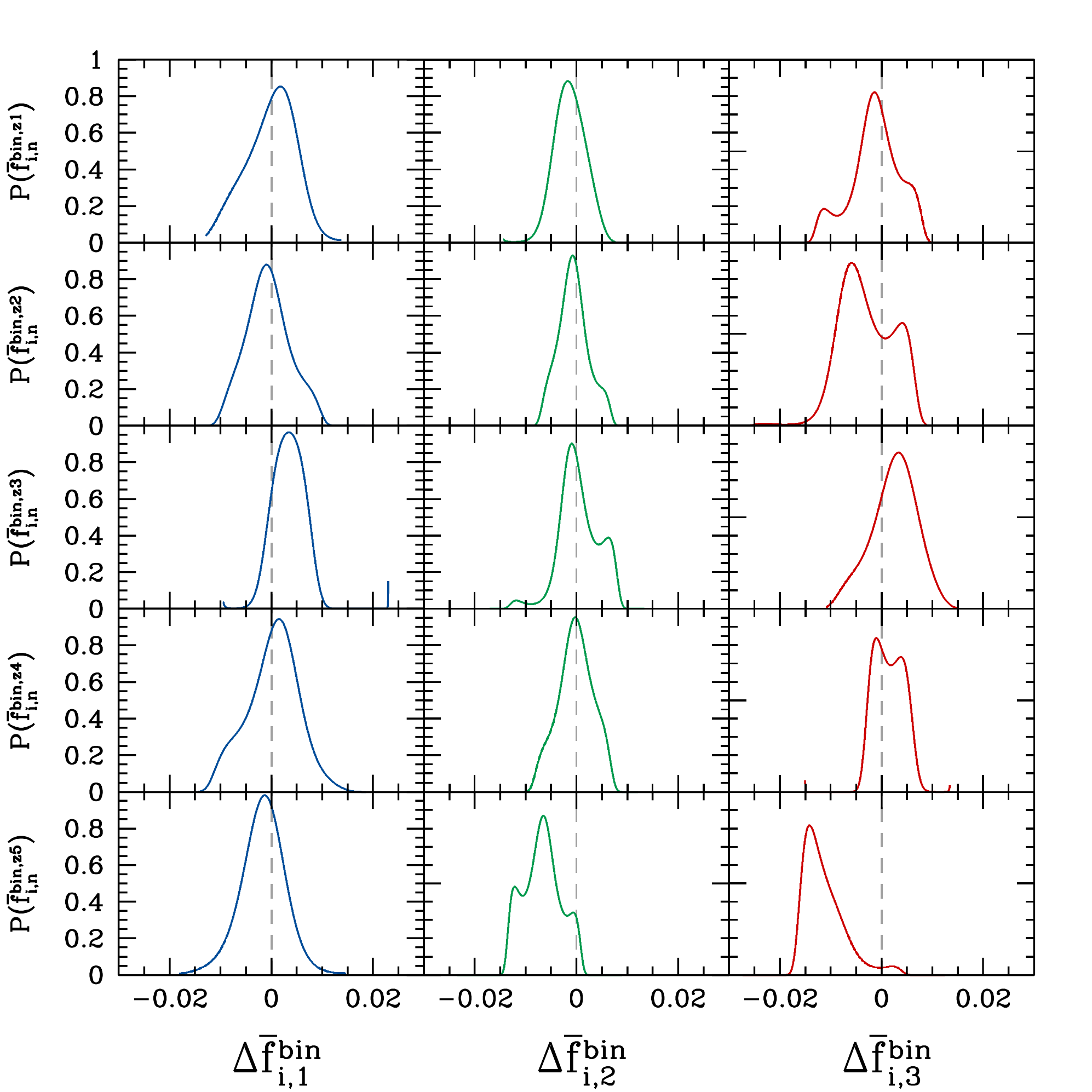}
\caption{{\it Left panel}: the constraint results of the cosmological parameters in different interloper fraction cases. The best-fits and 1$\sigma$ C.L. derived from the MCMC are shown as data points with error bars. We assume $f_{\rm i}(z)=f_{n}(1+z)$, and $f_n=f_1$, $f_2$, and $f_3$ denote the cases of the fiducial values $f^{\rm fid}_1$=0.01, $f^{\rm fid}_2$=0.05, and $f^{\rm fid}_3$=0.1. The first column shows the fitting result that $f_{\rm i}(z)$ in the five contaminated redshift intervals are known exactly (i.e. using fiducial or true $f_{\rm i}(z)$ in the constraint). The second column denotes the result that using the correctly derived or true $\bar{f}_{\rm i}^{\rm bin}$ to replace the true $f_{\rm i}(z)$ in the five contaminated redshift intervals. The third column shows the result for setting the five $\bar{f}_{\rm i}^{\rm bin}$ as free parameters in the MCMC with parameter range $(\bar{f}_{\rm i}^{\rm bin}-0.03, \bar{f}_{\rm i}^{\rm bin}+0.03)$. This range is determined based on $\Delta \bar{r}_{\rm t}\lesssim0.02$ derived from the simulations.  As comparison, the fourth column is the result without $f_{\rm i}$ considered in the fitting process. {\it Right panel}: the 1-d PDFs of $\Delta \bar{f}^{\rm bin}_{{\rm i},n}=\bar{f}^{\rm bin}_{{\rm i},n}-\bar{f}^{\rm bin,fid}_{{\rm i},n}$ in the five contaminated redshift intervals (from the top to bottom rows) for the free $\bar{f}^{\rm bin}_{\rm i}$ case, where $n=1$, 2, and 3 denote $f_n$=0.01, 0.05, and 0.1 cases, respectively.}
\label{fig:cons}
\end{figure}

The RSD galaxy power spectrum can be expanded in Legendre polynomials \citep{Ballinger96,Taylor96}
\be
P(k,\mu,z) = \sum_{\ell} P_{\ell}(k,z) \mathcal{L}_{\ell}(\mu),
\ee
where $\mathcal{L}_{\ell}(\mu)$ is the Legendre polynomials, and only the nonvanishing terms $\ell$=(0, 2, 4) are considered, and $P_{\ell}(k,z)$ is the multipole power spectrum. After including the Alcock-Paczynski effect \citep{Alcock79}, the galaxy multipole power spectrum is given by
\be
P_{\ell}(k,z) = \frac{2\ell+1}{2\alpha^2_{\perp}\alpha_{\parallel}} \int_{-1}^1 {\rm d}\mu P(k',\mu',z)\mathcal{L}_{\ell}(\mu),
\ee
where $\alpha_{\perp}(z)=D_{\rm A}(z)/D_{\rm A}^{\rm f}(z)$ and $\alpha_{\parallel}=H^{\rm f}(z)/H(z)$ are the scaling factors in the transverse and radial directions, respectively. $D_{\rm A}(z)$ is the angular diameter distance, and the superscript `f' means the quantities in the fiducial cosmology. $k'=\sqrt{k'^2_{\parallel}+k'^2_{\perp}}$ and $\mu'=k'_{\parallel}/k'$ are the apparent wavenumber and cosine of the angle along the line of sight, respectively, and $k'_{\perp}=k_{\perp}/\alpha_{\perp}$ and $k'_{\parallel}=k_{\parallel}/\alpha_{\parallel}$. The covariance matrix of the galaxy multipole power spectrum can be calculated by \citep[see e.g.][]{Taruya10}
\ba \label{eq:Del_Pk}
&{\rm Cov}&[P_{\ell,{\rm t}}(k,z),P_{\ell',{\rm t}}(k,z)] = \frac{(2\ell+1)(2\ell'+1)}{N_{\rm m}(k,z)} \nonumber\\
               &\times& \int_0^1 d\mu\, \mathcal{L}_{\ell}(\mu) \mathcal{L}_{\ell'}(\mu) P^2_{\rm t}(k,\mu,z),
\ea
where $N_{\rm m}$ is the number of modes in a wavenumber interval $\Delta k$, which can be estimated as $N_{\rm m}(k,z) = 2\pi k^2\Delta k\, V_{\rm s}(z)/(2\pi)^3$. Here $V_{\rm s}(z)$ is the survey volume at $z$. 
When generating the mock data of the CSST 3-d spectroscopic galaxy clustering survey, we divide the redshift range from $z=0$ to 1.2 into six intervals with $\Delta z=0.2$. Then we calculate $P_{\ell,{\rm t}}(k,z)$ and ${\rm Cov}[P_{\ell,{\rm t}}(k,z),P_{\ell',{\rm t}}(k,z)] $ for $\ell=0$, 2, and 4 at redshift around 0.1, 0.3, 0.5, 0.7, 0.9, and 1.1, respectively. We also add a random Gaussian distribution derived from the covariance matrix on each mock data point. 

After generating the mock data, we explore the constraints on the cosmological and other parameters. The Markov Chain Monte Carlo (MCMC) method is adopted to perform the constraint. We consider seven cosmological parameters (i.e. $\Omega_{\rm b}$, $\Omega_{\rm M}$, $w_0$, $w_a$, $h$, $\sigma_8$, and $n_{\rm s}$), the parameters of real galaxies in the six redshift intervals (i.e. six  $b_{\rm g}$, $\sigma_{\nu0}$, and $P_{\rm shot}$), and the parameters of interloper galaxies in five lower redshift intervals (i.e. five $b^{\rm i}_{\rm g}$ and $\sigma^{\rm i}_{\nu0}$). Flat priors of these parameters are assumed in the MCMC fitting process, and we have $\Omega_{\rm b}\in(0,0.1)$, $\Omega_{\rm M}\in(0,1)$, $w_0\in(-10,10)$, $w_a\in(-20, 20)$, $h\in(0,1)$, $\sigma_8\in(0.4,1)$, $n_{\rm s}\in(0.9,1.1)$, $b_{\rm g}$ or $b^{\rm i}_{\rm g}\in(0,4)$, $\sigma_{\nu0}$ or $\sigma^{\rm i}_{\nu0}\in(0,10)$, ${\rm log_{10}}P_{\rm shot}\in(0,4)$.  The Metropolis-Hastings algorithm is used to determine the accepted probability of a new chain point in the MCMC \citep{Metropolis53,Hastings70}. We run 16 chains and obtain about 100,000 points for each chain. After burn-in and thinning process, totally about 10,000 chain points are used to derive the probability distribution functions (PDFs) for all of the free parameters.

In the Left panel of Figure~\ref{fig:cons}, we show the constraint results of the seven cosmological parameters in different interloper fraction cases. The four columns show the fitting results of using the true values of  $f_{\rm i}(z)$ and $\bar{f}_{\rm i}^{\rm bin}$, free $\bar{f}_{\rm i}^{\rm bin}$, and without $f_{\rm i}$ considered in the five contaminated redshift intervals, respectively. We can find that, although it seems a bit worse, the results from the true and free $\bar{f}_{\rm i}^{\rm bin}$ are comparable to that from the true $f_{\rm i}(z)$. We can obtain similar result as the true $f_{\rm i}(z)$ case for $f_n\lesssim 0.01$, and mildly worse (with larger error) but acceptable result for $f_n\lesssim 0.1$. This proves the feasibility and validation of our method. In practice, we can adopt the method of setting the free $\bar{f}_{\rm i}^{\rm bin}$ in the MCMC with its parameter range derived from the angular correlation functions or power spectra. In the right panel, we show the 1-d PDFs of $\Delta \bar{f}^{\rm bin}_{\rm i}$ derived from the MCMC in the free $\bar{f}_{\rm i}^{\rm bin}$ case with different $f_n$ assumed. We can find that the correct $\bar{f}_{\rm i}^{\rm bin}$ in each redshift interval can be correctly derived at $z\lesssim1$, and the result will be better for smaller interloper fraction.

\section{discussion}

We notice that only H$\alpha$ and [OIII] lines are considered in above discussion, which are the two main emission lines that can be measured in the CSST spectroscopic survey. The [OII]$3727{\rm \AA}$ line is not the major concern here, since the number density of the [OII] galaxies observed by the CSST is relatively small. Most [OII] galaxies will be observed in the CSST $GU$ and the first half of $GV$ bands with low filter transmissions and detector quantum efficiencies, and they are mainly contaminated at high redshifts with $z\gtrsim0.7$ \citep{Zhou21}. But [OII] line can be more important in other spectroscopic surveys focusing on the near-infrared bands, such as $Euclid$ and RST. 

Besides, this method also can be applied to the intensity mapping surveys by changing the definition of the interloper fraction to be $f_{\rm i}(z)=\bar{I}_{\rm i}(z)/\bar{I}_{\rm t}(z)$, where $\bar{I}_{\rm i}$ and $\bar{I}_{\rm t}$ are the mean intensities of interloper and total galaxies, respectively. The contamination of interloper galaxies is a serious issue in the intensity mapping, especially for the surveys observing atomic and molecular emission lines \citep[see e.g.][]{Visbal10,Gong14,Silva15}. Since the intensity mapping does not resolve individual galaxies but measures the cumulative flux of all sources in a voxel with low spatial and spectral resolutions, the statistical method of eliminating the interloper contamination is an ideal choice. We will study it using this method in details in a future work.

\begin{acknowledgments}
YG and HTM acknowledge the support of NSFC-11822305, NSFC-11773031, NSFC-11633004, MOST-2018YFE0120800, 2020SKA0110402, and CAS Interdisciplinary Innovation Team. PJZ acknowledges the support of NSFC-11621303. XLC acknowledges the support of the National Natural Science Foundation of China through grant No. 11473044, 11973047, and the Chinese Academy of Science grants QYZDJ-SSW-SLH017, XDB 23040100, XDA15020200. This work is also supported by the science research grants from the China Manned Space Project with NO.CMS-CSST-2021-B01 and CMS-CSST-2021-A01.
\end{acknowledgments}


\end{document}